\documentclass[aps,nofootinbib,amsmath,prd,twocolumn,showpacs,superscriptaddress,groupedaddress]{revtex4-1}

\usepackage[american]{babel}
\usepackage{amsfonts}
\usepackage{amsmath}
\usepackage{amssymb}
\usepackage{slashed}
\usepackage{xcolor}
\usepackage{bm}
\usepackage{float}
\usepackage[utf8]{inputenc}
\usepackage[macrosonly]{chet}
\usepackage{nicefrac}
\usepackage{hyperref}
\usepackage{url}
\usepackage{array}
\usepackage{multirow}
\usepackage{comment}
\usepackage[normalem]{ulem}

\usepackage{mathtools,tensor,mhchem}

\usepackage{lmodern}
\usepackage{tikz}
\usetikzlibrary{decorations.pathmorphing}
\usetikzlibrary{positioning}

\tikzset{graviton/.style={decorate, decoration={snake, segment length=2mm, amplitude=0.6mm}}}

\newcommand{\dd}{{\rm d}}

\begin{document}

\title{General relativity and its modifications
as effective field theories of $GL(4)$ breaking}

\author{Tommaso Viacava}
\email{t.viacava@studenti.unipi.it}
\affiliation{Universit\`a di Pisa, Largo Bruno Pontecorvo 3, 56127 Pisa, Italy}

\author{Omar Zanusso}
\email{omar.zanusso@unipi.it}
\affiliation{Universit\`a di Pisa and INFN - Sezione di Pisa, Largo Bruno Pontecorvo 3, 56127 Pisa, Italy}

\begin{abstract}

We discuss General Relativity and some of its modifications as effective field theories of the breaking of the affine general linear group using the coset approach and the Maurer-Cartan form. While the ultimate goal is to break the symmetry down to the Lorentz subgroup, we argue that there are three interesting breaking cascades in which the affine general linear group is first broken to a larger subgroup that contains the Lorentz group. In all cases, the breaking of translations always generates Goldstone modes that can be interpreted as metric degrees of freedom. We argue that the effective field theories that emerge as a result of the different breaking patterns are all dynamically equivalent at the lowest order of the derivative expansion, e.g., using Einstein-Hilbert-like actions, but can differ beyond the leading order.

\end{abstract}

\maketitle

\section{Introduction}\label{sect:intro}
%

The study of General Relativity (GR) as an effective theory through the lenses of the Coset Construction and non-linear realisations is a long-standing approach \cite{CCWZ1}, starting from the seminal works by Volkov \cite{Volkov:1973ix} and Ogievetsky \cite{Ogievetsky:1974} on the spontaneous symmetry breaking of spacetime symmetries. Important recent attempts to formulate gravitational theories in this sense have been attempted by Tomboulis in Ref.~\cite{Tomboulis}, in which the building blocks of GR have been derived from the spontaneous symmetry breaking of the four-dimensional affine general linear group $G_A(4,\mathbb{R}) \equiv GL(4,\mathbb{R}) \ltimes \mathbb{R}^4$ to the Lorentz group $SO(3,1)$, and by Sarmiento and Rodrìguez in Ref.~\cite{sarmiento2025brokenweylgravitydynamical}, in which the Weyl phase of a theory possessing an $\operatorname{ISO}(3,1) \ltimes \mathbb{R}_{{\rm dil}}$ symmetry is spontaneously broken down to the Lorentz group $SO(3,1)$.

The aim of this work is to generalize the results of Refs.~\cite{Tomboulis,sarmiento2025brokenweylgravitydynamical} by providing a systematic classification of gravitational theories that are based on different symmetry breaking patterns. Specifically, we study the effective field theories (EFTs) derived from the spontaneous symmetry breaking of a global $G_A(4,\mathbb{R})$ as in Ref.~\cite{Tomboulis} to its various subgroups, selecting the broken generators among those of the Lie algebra $\mathfrak{g}_A(4,\mathbb{R})$.
The algebra consists of the following commutation relations 
\begin{equation}
	\label{GLalgebra}
	\begin{split}
		[P^a, P^b] &= 0 \,, \\
		[G^{a}{}_{b}, P^c] &= i \delta^c_b P^a \,,\\
        [G^{a}{}_{b}, G^{c}{}_{d}] &= i(\delta^c_b G^a_d - \delta^a_d G^c_b ) \,.
	\end{split}
\end{equation}
Introducing the ``metric'' $\eta_{ab}={\rm diag}(-,+,+,+)$ and defining $G_{ab}=\eta_{ac}G^c{}_b$,
we can separate the twenty generators of $\mathfrak{g}_A(4,\mathbb{R})$ as 
\begin{itemize}
    \item $4$ translations $P^a$,
    \item $10$ symmetric generators $T_{ab} = 2G_{(ab)}$, which can be decomposed into $9$ traceless generators $S_{ab} = T_{ab}-\frac{1}{4} \eta_{ab} D$ and one scalar generator $D= T^a{}_a$ corresponding to dilatations,
    \item $6$ Lorentz generators $J_{ab}= 2G_{[ab]}$,
\end{itemize}
with commutation relations provided in the appendix.

The available options in selecting a breaking pattern of $G_A(4,\mathbb{R})$ to a subgroup are determined by which of the symmetric generators $T_{ab}$ are broken. In fact, the translations must always be broken in order for the Maurer--Cartan form to contain a vierbein, which is the fundamental tool for recovering a metric in the gravitational EFT, while the Lorentz generators $J_{ab}$ must always remain unbroken to preserve Lorentz invariance. In this work, we begin by reviewing the approach of Ref.~\cite{Tomboulis} in which all $T_{ab}$ are broken, and then study three alternative scenarios classified by the breaking of either $S^{ab}$ or $D$. We will argue that, in order to recover a well-defined EFT based on the metric tensor in the alternative scenarios, an additional \emph{gauge}-breaking pattern down to the Lorentz group is necessary and that the resulting EFTs can be interpreted as well-known alternative formulations of GR.
Notice that throughout this work we have chosen to work in $d=4$ spacetime dimensions, but a generalization to arbitrary $d$ is trivial. In the following, we will denote the coset space as $G/H$, where $G$ has always to be intended as $G_A(4,\mathbb{R})$, while $H$ will change depending on which breaking pattern we will study.

\section{Breaking to the Lorentz Group}\label{sect:Lorentz}

We begin by reviewing the coset construction associated with the symmetry breaking pattern
\begin{equation}
\label{eq:breaking}
    GL(4,\mathbb{R}) \ltimes \mathbb{R}^4 \, \longrightarrow \, SO(3,1) \,.
\end{equation}
This is the pattern investigated in Ref.~\cite{Tomboulis} that also serves as a foundation to illustrate the general procedure. A generic coset representative for the quotient space of Eq.~\eqref{eq:breaking} can be parametrized as
\begin{equation}
\label{eq:coset_rep}
\Gamma(x, \xi) = \exp(i x^a P_a) \exp \bigg( \frac{i}{2} \xi_{ab}(x) T^{ab} \bigg) \,,
\end{equation}
where $\xi_{ab}(x)$ are the Goldstone fields associated with the broken generators $T^{ab}$. From Eq.~\eqref{eq:coset_rep}, we can compute the Maurer--Cartan form
\begin{equation}
    \Omega = \Gamma^{-1}\dd\Gamma \,,
\end{equation}
and expand it in the algebra basis as
\begin{equation}
	\Omega = i \hat{\omega}^a P_a + \frac{i}{2} D_{ab} T^{ab} + \frac{i}{2}\omega_{ab} J^{ab} \,.
\end{equation}
A lengthy but straightforward algebraic manipulation yields the following explicit expressions for the components:
\begin{equation}
\label{eq:MC_components}
	\begin{cases}
		\hat{\omega}^a &= e^a{}_{\mu} \dd x^\mu \,, \\
		D_{ab} &= -\frac{1}{2}\{e^{-1}, \dd e\}_{ab} \,,\\
		\omega_{ab} &= -\frac{1}{2}[e^{-1}, \dd e]_{ab} \,,
	\end{cases}
\end{equation}
where we have introduced
\begin{equation}
\label{eq:vielbeinLorentz}
	e^{a}{}_{\mu} = \exp(-\xi)^a{}_{\mu} \,.
\end{equation}
The components in Eq.~\eqref{eq:MC_components} have the following geometric interpretations:
\begin{itemize}
    \item The $1$-form $\hat{\omega}^a$ is an anholonomic basis on the cotangent space, thus $e^a{}_{\mu}$ is a vierbein (or tetrad). Globally, the component of the Maurer--Cartan form along the broken translations acts as a \textit{soldering form}. We denote $e^\mu{}_a$ the inverse of the vierbein matrix.
    \item The field $D_{ab}$ acts as the \textit{covariant derivative} of the Goldstone fields. For instance, at linear order in the $\xi$-expansion, we recover
    \begin{equation}
        D_{ab} = -\frac{1}{2}\{e^{-1}, \dd e\}_{ab} = \dd\xi_{ab} + \mathcal{O}(\xi^2) \,.
    \end{equation}
    Its transformation behavior under the action of the full group $G$ ensures that these derivative terms transform covariantly, thus justifying the name. 
    \item Finally, $\omega_{ab}$ has the transformation properties required to construct a covariant derivative for matter fields, i.e., it provides a (preliminary) \emph{spin-connection} $\omega_{abc}$ which is defined as
    \begin{equation}
        \omega_{ab} \equiv \hat{\omega}^c \omega_{cab} \,.
    \end{equation}
\end{itemize}

We concentrate first on the spin-connection.
Let $\psi$ denote a field that transforms under an irreducible representation ${\cal R}$ of the Lorentz subgroup as
\begin{equation}
    \psi \longrightarrow \psi' = {\cal R}\left( \exp\left[\frac{i}{2} u_{ab}(x) J^{ab}\right] \right) \cdot \psi \,.
\end{equation}
We can construct a covariant derivative as
\begin{equation}
\label{eq:cov_der}
	D_c \psi(x) \equiv e^\mu{}_{c} \partial_\mu \psi(x) + \frac{i}{2} \omega_{cab}{\cal R}(J^{ab})\cdot \psi(x) \,.
\end{equation}
It is worth noting that the spin connection is not uniquely determined, as already pointed out in Ref.~\cite{Tomboulis}. We can always modify it by adding covariantly transforming terms that preserve the antisymmetry in the last two indices. By including suitable combinations of $D_{cab}$ terms, we obtain a generalized connection
\begin{equation}
	\label{eq:general_connection}
	\Omega_{cab} = \omega_{cab} + c_1(D_{bac} - D_{abc}) \,,
\end{equation}
where $c_1$ is an arbitrary coefficient\footnote{In principle, one could leave an arbitrary coefficient $c_0$ in front of $\omega_{cab}$ in Eq.~\eqref{eq:general_connection}. However, requiring standard local Lorentz gauge covariance strictly fixes this coefficient to $c_0 = 1$.}.

With the building blocks of Eq.~\eqref{eq:MC_components} at hand, we can construct Lorentz-invariant quantities that are, by virtue of the coset construction, also automatically invariant under the full broken affine group. Algebraically, this is achieved via the \textit{dressing} procedure \cite{CCWZ1, CCWZ2}, which lifts a representation of the unbroken subgroup to a representation of the full group through a non-linear dressing function of the Goldstone bosons. In the case under investigation, this function is precisely the inverse vierbein $e^\mu{}_{a}$. For example, given a Lorentz covector $v_a$ that transforms as
\begin{equation}
    v_a \longrightarrow v'_a = (\Lambda^{-1})_a{}^{b} v_b \,,
\end{equation}
it can be shown that
\begin{equation}
\label{eq:dressing}
    V_\mu = e^a{}_\mu v_a
\end{equation}
is a true covector under the full group.\footnote{Indeed, given the transformation rules for the vielbein $e\longrightarrow e'=D(g)\; e \;D(h^{-1})(\xi,g)$, we can consider a Lorentz vector $v \longrightarrow v' = D(h(\xi,g) v$, with $D$ being the representation of the group $G$. If we choose $D$ as the fundamental-defining representation, when restricted to $H$ it reduces to the fundamental representation in Lorentz, thus allowing the cancellation of the factors and the correct transformation $V\longrightarrow V' = D(g) V$.} 

Clearly, Eq.~\eqref{eq:dressing} generalizes to any $SO(3,1)$-tensor of integer spin, provided that the finite-dimensional representation of the full group $G$ contains the irreducible representation of the unbroken subgroup $H$ to be lifted, as discussed in Ref.~\cite{CCWZ2}. According to a fundamental theorem by Cartan \cite{Cartan:1981}, there are no finite-dimensional spinor representations of $GL(4,\mathbb{R})$. As a consequence, the dressing procedure cannot be applied to half-integer spin fields, and fermions must strictly reside on the Lorentz fiber, coupling exclusively through the spin-connection. We will briefly return to the implications of this feature in the final discussion.

In geometric terms, the vierbein defines spacetime tensors starting from Lorentz ones, which reside in the abstract vector space associated with the $SO(3,1)$ principal bundle. Among the possible $SO(3,1)$ tensors, of particular relevance is the Minkowski metric tensor, which allows us to construct a spacetime metric via the aforementioned dressing procedure:
\begin{equation}
\label{eq: spacetime metric}
    g_{\mu\nu} = e^a{}_{\mu} e^b{}_{\nu} \eta_{ab} \,.
\end{equation}    
In the analysis of alternative breaking patterns of the next sections, Eq.~\eqref{eq: spacetime metric} together with the specific form of the vierbein will establish a classification of possible gravitational theories.

Having established the geometric ingredients from the Maurer--Cartan form, one can compute the torsion and curvature $2$-forms (in Lorentz indices) via Cartan's structure equations:
\begin{equation}
\label{eq:Cartan structure}
	\begin{split}
		& T^a = \dd \hat{\omega}^a + \Omega^a{}_{b} \wedge \hat{\omega}^b,\\
		& R^a{}_{b} = \dd\Omega^a{}_{b} + \Omega^a{}_{c} \wedge \Omega^c{}_{b} \,.
	\end{split}
\end{equation}
The vierbein components then map these geometric structures onto the spacetime manifold. The explicit expressions for the curvature and torsion are given in terms of the Goldstone fields $\xi$ via Eq.~\eqref{eq:MC_components} and the choice of the generalized connection in Eq.~\eqref{eq:general_connection}. 

In particular, it can be proven that the unique choice of the coefficient $c_1$ that forces the torsion to vanish identically is $c_1 = -1$. This specific parameter tuning recovers the standard geometric building blocks of General Relativity, as assumed in Ref.~\cite{Tomboulis}, while, for arbitrary coefficients, the torsion does not vanish and the resulting Effective Field Theory (EFT) must incorporate it alongside the curvature. Nevertheless, the true nature of the resulting theory must be investigated through the construction of an effective action.

\section{Alternative breaking patterns}

The procedure outlined previously can be carried along to study alternative breaking patterns from $G_A(4, \mathbb{R})$ to any subgroup $H$. In doing this, we must pay attention to the possibility of obtaining a spacetime metric tensor as in Eq.~\eqref{eq: spacetime metric}. For instance, tracing back the steps of Sect.~\ref{sect:Lorentz}, this can be achieved by soldering a generic fiber metric $\gamma_{ab}$ to the spacetime manifold,
\begin{equation}
\label{eq: generic metric}
    g_{\mu\nu} = e^a{}_{\mu} e^b{}_{\nu} \gamma_{ab} \,,
\end{equation}
where $\gamma$ must be an invariant tensor under the unbroken group $H$. In the following sections, we explore the cases
\begin{equation*}
    \begin{split}
        &H = SL(4,\mathbb{R}) \,,\\
        &H = SO(3,1) \ltimes \mathbb{R}_{{\rm dil}} \,,
    \end{split}
\end{equation*}
with $\mathbb{R}_{{\rm dil}}$ denoting the Abelian subgroup of dilatations generated by $D$. In both these scenarios, we cannot define an invariant rank-$2$ symmetric tensor; thus, in order to obtain an invariant spacetime metric, we need to postulate an additional breaking pattern down to the Lorentz group, resulting in a \textit{symmetry breaking cascade} of the type
\begin{equation}
    G_A(4,\mathbb{R}) \, \longrightarrow \, H \, \longrightarrow \, SO(3,1) \,.
\end{equation}

In this way, one can obtain an invariant spacetime metric via Eq.~\eqref{eq: generic metric} with $\gamma \equiv \eta$ being the Minkowski tensor. However, in performing such an operation, we need to pay attention to a remarkable feature of the coset construction: after the first breaking $H$ is a local symmetry group. This affects the procedure on two fronts. On the one hand, the resulting Goldstone bosons are not physical degrees of freedom of the theory and can always be eliminated by means of a unitary gauge. On the other hand, the Maurer--Cartan form also needs to be modified in order to be covariant under the local symmetry group. In general, this can be accomplished by introducing an additional gauge field $A$ \cite{Delacretaz2014}, which defines a covariant Maurer--Cartan form as
\begin{equation}
\label{eq:Mod_CC}
    \Omega \equiv \Gamma^{-1}\nabla\Gamma = \Gamma^{-1}\dd\Gamma + \Gamma^{-1} A^I V_I \Gamma \,, 
\end{equation}
with $V_I$ being the gauged generators of the group $H$. We will follow this strategy in Section~\ref{sect:dilatations} when gauging the dilatations in one of the alternative breaking patterns.

However, the symmetry breaking cascade suggests an alternative way to perform this construction without requiring the addition of an external gauge field. Without going into the mathematical details\footnote{This is a direct consequence of a standard result in fiber bundle theory known as the \textit{Bundle structure theorem} \cite{Steenrod:1951} ($\S\ 7.4$).}, we observe that, given a symmetry breaking cascade
\begin{equation}
    G \longrightarrow H_1 \longrightarrow H_2,
\end{equation}
it is always possible to choose a coset representative $\tilde{\Gamma} \in G/H_2$ as
\begin{equation}
    \tilde{\Gamma} = \Gamma_1 \cdot \Gamma_2,
\end{equation}
with $\Gamma_1 \in G/H_1$ and $\Gamma_2 \in H_1/H_2$. Its associated Maurer--Cartan form reads
\begin{equation}
\label{eq:MC Cascade}
    \Omega \equiv \tilde{\Gamma}^{-1}\dd\tilde{\Gamma} = \Gamma_2^{-1}\Omega_1 \Gamma_2 + \Omega_2,
\end{equation}
and it is straightforward to prove that it satisfies the expected transformation law under $G$:
\begin{equation}
    \Omega' = h_2 \Omega h_2^{-1} + h_2 \dd h_2^{-1}.
\end{equation}

\subsection{Breaking of dilatations}

The first breaking pattern we analyze consists in breaking only the dilatations $D$, i.e.,
\begin{equation}
    G_A(4,\mathbb{R}) \, \longrightarrow \, SL(4,\mathbb{R})\,,
\end{equation}
with $SL(4,\mathbb{R})$ being the $4$-dimensional special linear group of volume-preserving transformations generated by the $S_{ab}$. We write the coset representative as
\begin{equation}
	\Gamma(x, \lambda) = \exp (i P \cdot x) \exp (i\lambda(x) D)\,,
\end{equation}
where $\lambda(x)$ denotes the scalar Goldstone boson related to the breaking of dilatations, which we refer to as the \textit{dilaton}. The associated Maurer--Cartan form reads
\begin{equation}
	\Gamma^{-1}\dd\Gamma = i\hat{\omega}^a P_a + i\hat{\omega}_\lambda D + \frac{i}{2}\omega_S^{ab} S_{ab} + \frac{i}{2}\omega_J^{ab} J_{ab} \,.
\end{equation}
An analogous calculation to the that leads to Eq.~\eqref{eq:MC_components} yields 
\begin{equation}
    \label{eq:Unimodular MC}
	\begin{cases}
		\hat{\omega}^a &= \exp(-\lambda(x)) \delta^a{}_{\mu} \dd x^\mu \,, \\
		\hat{\omega}_\lambda &= \dd\lambda(x) \,,\\
		\omega_{S} &= 0 \,, \\
        \omega_J &= 0 \,.
	\end{cases}
\end{equation}
Note that in this case the algebra does not give any component along both the Lorentz $J$ and the shear-traceless generators $S$.

It should be evident by now that this breaking pattern, unlike the previous one, does not provide a nonvanishing connection on the induced principal $SL(4,\mathbb{R})$ bundle, thus resulting in a less rich geometric structure. Furthermore, as discussed previously, we do not have an invariant metric tensor. There is, however, an $SL(4,\mathbb{R})$ invariant does exist, the volume form, which turns out to be 
\begin{equation}
	{\rm d} V = \exp(-4\lambda(x)) \dd x^1 \wedge \dd x^2 \wedge \dd x^3 \wedge \dd x^4 \,.
\end{equation}
Note that the volume form is not fixed, but all its dynamics is dictated by the dilaton. In order to obtain a viable gravitational theory, as discussed above, we need to postulate an additional breaking pattern,
\begin{equation}
    SL(4,\mathbb{R}) \, \longrightarrow \, SO(3,1) \,,
\end{equation}
which breaks also the traceless shear generators $S^{ab}$. According to the previous discussion, the Maurer--Cartan form associated with the breaking cascade is given by Eq.~\eqref{eq:MC Cascade} with
\begin{equation}
    \Gamma_2 = \exp\bigg(\frac{i}{2}\sigma_{ab}(x) S^{ab} \bigg)
\end{equation}
and $\Omega_1$ given by its components in Eq.~\eqref{eq:Unimodular MC}. In this case the first piece in eq. \eqref{eq:MC Cascade} reads
\begin{equation}
\begin{split}
    \Gamma_2^{-1} \Omega_1 \Gamma_2 &= i \, e^{-\lambda} \big(\Gamma_2^{-1} P_a \Gamma_2\big) dx^a + i \, \dd\lambda \, D \\
    &=i \, e^{-\lambda} \big(e^{-\sigma}\big)^a{}_b dx^b P_a + i \, \dd\lambda \, D 
\end{split}
\end{equation}
while the second piece gives the decomposition along $\mathfrak{sl}(4,\mathbb{R})$ and Lorentz
\begin{equation}
    \Omega_2 = \Gamma_2^{-1}\dd\Gamma_2 = \frac{i}{2} D_{ab} S^{ab} + \frac{i}{2} \omega_{ab} J^{ab}.
\end{equation}

A straightforward calculation then gives
\begin{equation}
    \label{eq:Unimodular cascade}
	\begin{cases}
		\hat{\omega}^a &= (\exp(-\sigma(x)))^a_{\phantom{a}b} \exp(-\lambda(x)) \delta^b_{\phantom{b} \mu} \dd x^\mu \,, \\
		\hat{\omega}_\lambda &= \dd\lambda(x) \,,\\
		\omega_{SO(3,1)} &= -\frac{1}{2}[\tilde{e}^{-1}, \dd\tilde{e}]_{ab} \,, \\
        D_{ab} &= -\frac{1}{2}\{\tilde{e}^{-1}, \dd\tilde{e}\}_{ab} \,,
	\end{cases}
\end{equation}
where $\tilde{e}$ is given by Eq.~\eqref{eq:vielbeinLorentz} with the substitution $\xi \rightarrow \sigma$, with the important difference that $\sigma$, unlike $\xi$, is an unphysical field and can be removed from the theory by a suitable gauge transformation. In order to investigate this possibility, we need to check the transformation rule for the coset representative $\Gamma_2$ under a gauge $SL(4,\mathbb{R})$ transformation, given by
\begin{equation}
\label{ex}
    \Gamma_2' = h_1(x) \Gamma_2 h_2^{-1}(h_1, \sigma)\,,
\end{equation}
where $h_1 \in SL(4,\mathbb{R})$ and $h_2 \in SO(3,1)$ is the Lorentz compensating factor to be determined in order for $\Gamma_2'$ to remain in the coset $H_1/H_2$. While this compensating factor cannot be determined in general in closed form, we can deduce it by means of infinitesimal transformations, as discussed in detail in Ref.~\cite[$\S\ 7.3.1$]{Brauner:2024juy}. Choosing 
\begin{equation}
\label{eq:gaugechoice}
    h_1(x) = \exp\bigg(\frac{i}{2}\epsilon_{ab}(x) S^{ab}\bigg), \quad h_2 = \exp\bigg(\frac{i}{2}\Theta_{ab}(x) J^{ab}\bigg)\,,
\end{equation}
with gauge parameter $\epsilon_{ab}$, we can compare the infinitesimal variation $\delta \Gamma_2 = \Gamma_2' - \Gamma_2$ with the first-order expansion in the gauge parameter and in $\Theta$ of Eq.~\eqref{eq:gaugechoice}, thus obtaining
\begin{equation}
\label{eq:Sigma}
    \sigma'_{ab}(x) = \sigma_{ab}(x) + \epsilon_{ab}(x) + \mathcal{O}(\epsilon^2) \,.
\end{equation}
The unitary gauge thus corresponds to $\epsilon_{ab}(x) = -\sigma_{ab}(x)$. In this gauge $\sigma'_{ab} = 0$, yielding a vanishing spin connection $\omega_{SO(3,1)}$ (as well as $D_{ab} = 0$) and consequently a vanishing curvature $2$-form. In this sense, we can think of the unitary gauge as a ``Weitzenb{\"o}ck'' one \cite{BeltranJimenez:2019esp}$\S 3.1$.

All the geometry is thus encoded in the torsion $2$-form $T^a = \dd e^a$ and the conformally flat metric:
\begin{equation}
\label{eq:UDmetric}
    g_{\mu\nu} = \exp(-2\lambda(x))\eta_{\mu\nu} \,.
\end{equation}
We note that this metric tensor has a unimodular splitting of the type
\begin{equation}
\label{eq:UDgeneral}
    g_{\mu\nu} = \exp(\chi) \bar{g}_{\mu\nu} \,,
\end{equation}
where $\bar{g}_{\mu\nu} = \eta_{\mu\nu}$ is unimodular and $\chi(x) = -2\lambda(x)$ is a scalar field which we can interpret as the Goldstone dilaton field. The structure in Eq.~\eqref{eq:UDgeneral} is typical of Unimodular-Dilaton (UD) theories of gravity \cite{Gielen:2018}. Note, however, that in this case all the spacetime dynamics is governed by the dilaton, which also enters the volume form and defines a non-vanishing torsion form.

From the general curvature $2$-form in Eq.~\eqref{eq:Cartan structure}, we can construct the Levi-Civita curvature directly from the metric tensor in Eq.~\eqref{eq:UDmetric}:
\begin{equation}
    \mathring{R} = 6\exp(2\lambda(x))\big(\square\lambda(x) - \partial_\mu\lambda(x)\partial^\mu\lambda(x)\big) \,,
\end{equation}
from which we can construct the standard Einstein--Hilbert action:
\begin{equation}
\label{eq:EH}
    S_{EH} = \frac{1}{2\kappa^2}\int \dd^{4}x \sqrt{-g}\mathring{R} \,.
\end{equation}
It is interesting to note that by a suitable redefinition of the scalar field:
\begin{equation}
    \phi(x) = \sqrt{\frac{6}{\kappa^2}} \exp(-\lambda(x)) \,,
\end{equation}
Eq.~\eqref{eq:EH} reads
\begin{equation}
    S_{EH} = \frac{1}{2}\int \dd^4x \; \eta^{\mu\nu}\partial_\mu \phi(x) \partial_\nu \phi(x) \,,
\end{equation}
which is precisely the expected action of a NLSM for a Goldstone boson $\phi$. Adding a cosmological constant term, we obtain the standard $\phi^4$ action with a coupling determined by the cosmological constant $\Lambda$:
\begin{equation}
    \lambda_0 = \frac{2}{3}\Lambda \kappa^2 \,.
\end{equation}

To sum up, the Einstein--Hilbert action with a cosmological constant term reproduces a standard $\phi^4$ theory in a flat spacetime. Solving the associated equation of motion gives the dynamics (at leading order in the derivative expansion) of the conformal factor appearing in Eq.~\eqref{eq:UDmetric}, which in principle can be used in cosmological models to describe the evolution of the universe.

\subsection{Unimodular Gravity}

The second breaking pattern that we study consists in breaking the traceless shear generators $S_{ab}$
\begin{equation}
    G_A(4,\mathbb{R}) \longrightarrow SO(3,1) \ltimes \mathbb{R}_{{\rm dil}} \,.
\end{equation}
The coset representative can be chosen to be
\begin{equation}
    \Gamma(x,\sigma) = \exp(i P \cdot x)\exp\bigg(\frac{i}{2}\sigma_{ab}(x) S^{ab}\bigg),
\end{equation}
with $\sigma_{ab}(x)$ being the traceless Goldstone associated to the broken traceless shear generators $S_{ab}$. In this case the Maurer-Cartan form reads as usual as
\begin{equation}
    \Gamma^{-1}\dd\Gamma = i\hat{\omega}^a P_a + i\hat{\omega}_\lambda D + \frac{i}{2}\omega_S^{ab} S_{ab} + \frac{i}{2}\omega_J^{ab} J_{ab},
\end{equation}
where in this case the Coset Construction yields to
\begin{equation}
\label{eq:MC_components_UD}
	\begin{cases}
		\hat{\omega}^a &= \exp(-\sigma)^a_{\phantom{a}\mu} \dd x^\mu \,,\\
		D_{ab} &= -\frac{1}{2}\{e^{\sigma}, \dd e^{-\sigma}\}_{ab} \,,\\
		\omega_{ab} &= -\frac{1}{2}[e^{\sigma}, \dd e^{-\sigma}]_{ab} \,.
	\end{cases}
\end{equation}
Under an $H$-transformation, the metric is no longer invariant, but transforms with a conformal factor coming from the $\mathbb{R}_{{\rm dil}}$ part of the symmetry group,
\begin{equation}
    g_{\mu\nu} \longrightarrow g'_{\mu\nu} = \exp(2\Omega(x)) g_{\mu\nu} \,,
\end{equation}
thus defining a Weyl geometry on the spacetime manifold, with a spacetime metric that is strictly unimodular, i.e., $|\det(g)| = 1$.

The consequences of this Weyl geometry will be explored in the next section, where the gauging of dilatations introduces a background gauge field that can be interpreted as the Weyl vector. Here, we simply assume the existence of an additional breaking pattern involving only dilatations that yields
\begin{equation}
\label{eq:MC_components_UD_full}
	\begin{cases}
		\hat{\omega}^a &= \exp(-\lambda(x)) \exp(-\sigma)^a_{\phantom{a}\mu} \dd x^\mu
        \,,\\
		\hat{\omega}_\lambda &= \partial_\mu \lambda(x) \dd x^\mu
        \,,\\
		D_{ab} &= -\frac{1}{2}\{e^{\sigma}, \dd e^{-\sigma}\}_{ab}
        \,,\\
		\omega_{ab} &= -\frac{1}{2}[e^{\sigma}, \dd e^{-\sigma}]_{ab}
        \,.
	\end{cases}
\end{equation}
Following the same procedure that has led to Eq.~\eqref{eq:Sigma} in the previous case, it is easy to show that, in this case, the Lorentz compensating factor is strictly the identity, while the dilaton transforms under gauge $\mathbb{R}_{{\rm dil}}$ transformations by a shift:
\begin{equation}
\label{eq:Tranform_dil}
    \lambda(x) \longrightarrow \lambda'(x) = \lambda(x) + \alpha(x) \,,
\end{equation}
with $\alpha(x) = -\lambda(x)$ corresponding to the unitary gauge. In this gauge, the spacetime metric tensor is given by
\begin{equation}
\label{eq:U_metric}
    g_{\mu\nu} = \exp(-\sigma)^a_{\phantom{a}\mu} \exp(-\sigma)^b_{\phantom{b}\nu} \eta_{ab} \,,
\end{equation}
which is invariant under $SO(3,1)$ transformations and has unit determinant. Consequently, we can conclude that the resulting theory corresponds to \textit{Unimodular Gravity}, where the spacetime dynamics is governed by a spin-$2$ traceless tensor.

\subsection{Weyl Gravity}\label{sect:dilatations}

The last pattern is a slight modification of the previous one, where the dilatations group $\mathbb{R}_{{\rm dil}}$ is now a gauged subgroup of the original general linear symmetry group. Due to this, we need to perform the modified version of the coset construction given in Eq.~\eqref{eq:Mod_CC},
\begin{equation}
\Omega \equiv \Gamma^{-1}\nabla \Gamma = \Gamma^{-1}\dd\Gamma + i \Gamma^{-1} b D \Gamma \,,
\end{equation}
where $b = b_\mu \dd x^\mu$ is an additional gauge field corresponding to the \textit{Weyl vector} (the Abelian gauge field associated to dilatations).
The components of the Maurer--Cartan form now read
\begin{equation}
\label{eq:MC_components_Weyl}
	\begin{cases}
		\hat{\omega}^a &= \exp(-\sigma)^a_{\phantom{a}\nu}(\delta^\nu_{\phantom{\nu}\mu} + b_\mu x^\nu) \dd x^\mu 
        \,,\\
		\hat{\omega}_\lambda &= b_\mu \dd x^\mu 
        \,,\\
		D_{ab} &= -\frac{1}{2}\{e^{\sigma}, \dd e^{-\sigma}\}_{ab} 
        \,,\\
		\omega_{ab} &= -\frac{1}{2}[e^{\sigma}, \dd e^{-\sigma}]_{ab}
        \,.
	\end{cases}
\end{equation}
The presence of $b_\mu$ results in a modification of the vierbein with respect to the previous pattern, which leads to the spacetime metric
\begin{equation}
\label{eq:Metric_Weyl}
    g_{\mu\nu} = \tilde{g}_{\mu\nu} + b_\nu x^\gamma \tilde{g}_{\mu\gamma} + b_\mu x^\gamma \tilde{g}_{\gamma\nu} + b_\mu b_\nu x^\rho x^\gamma \tilde{g}_{\rho\gamma} \,,
\end{equation}
where $\tilde{g}$ denotes the unimodular metric in Eq.~\eqref{eq:U_metric}. The explicit dependence of the metric tensor in Eq.~\eqref{eq:Metric_Weyl} on the spacetime coordinates is a consequence of translations being an ideal of the Lie algebra, specifically due to the nontrivial commutator $[P, D] \propto P$.

In physical terms, the Weyl vector couples to the Goldstone bosons associated with the breaking of translations, which are the coordinates themselves. This peculiarity breaks the unimodularity of the metric, in fact
\begin{equation}
\label{eq:Hyp}
    |\det(g)| = (1 + b_\mu x^\mu)^2 \,.
\end{equation}
Furthermore, Eq.~\eqref{eq:Hyp} defines a hypersurface on which the metric determinant vanishes, making the vierbein non-invertible. We can think of this hypersurface as the boundary of the local coordinate patch where the exponential parametrization acts as a diffeomorphism. 

Another salient feature of this pattern is the appearance of a nonmetricity tensor in the theory. To see this, let us consider the covariant derivative of a generic $H$-tensor $T$,
\begin{equation}
	\label{eq:cov_tensor}
    \begin{split}
        D_{c}T^A = &(e^{-1})_c{}^{\mu} \partial_{\mu}T^A + \frac{i}{2}[\Omega_{c a b} J^{a b}]^A{}_{B} \, T^{B} \\
         &+ i(e^{-1})_{c}{}^{\mu} b_{\mu} [D]^A{}_{B} \, T^{B}
         \,,
    \end{split}
\end{equation}
where
capitalized Latin indices
denote the indices of the $H$-representation under which the tensor transforms. We note that, apart from the usual spin connection term, in this case there is an extra term proportional to the Weyl vector; it is easy to see that, if $T$ has Weyl weight $w$,\footnote{%
The weight is the charge associated to the Abelian potential, i.e., the negative of the mass dimension.
} the generators of dilatations in its representation read
\begin{equation}
    [D]^A{}_{B} = -iw \delta^A{}_{B} \,.
\end{equation}
By considering the covariant derivative of the metric tensor,
\begin{equation}
    \nabla_{\rho}g_{\mu\nu} = e^a{}_{\mu} e^b{}_{\nu} e^c{}_{\rho} D_{c}\eta_{ab} \,,
\end{equation}
so we can prove that
\begin{equation}
    \nabla_\rho g_{\mu\nu} = 2b_\rho g_{\mu\nu} \neq 0 \,,
\end{equation}
since $\eta_{ab}$ is a rank-$2$ $H$-tensor with weight $w = 2$ \cite{Codello:2012sn,Sauro:2022chz,Zanusso:2023vkn}.

Choosing the coefficients in the spin connection $\Omega$ such that the torsion vanishes, we can compute the spacetime connection associated with this geometry, which, according to the fundamental theorem of Riemannian geometry, can be decomposed as
\begin{equation}
    \Gamma_{\mu \phantom{\alpha} \nu}^{\phantom{\mu} \alpha} = \mathring{\Gamma}_{\mu \phantom{\alpha} \nu}^{\phantom{\mu} \alpha} + L_{\mu \phantom{\alpha} \nu}^{\phantom{\mu} \alpha} \,,
\end{equation}
with $\mathring{\Gamma}$ being the Levi--Civita connection and $L$ the disformation tensor. A short computation yields
\begin{equation}
    \Gamma_{\mu \phantom{\alpha} \nu}^{\phantom{\mu} \alpha} = \mathring{\Gamma}_{\mu \phantom{\alpha} \nu}^{\phantom{\mu} \alpha} - b_\mu \delta^\alpha_{\phantom{\alpha}\nu} - b_\nu \delta^\alpha_{\phantom{\alpha}\mu} + b^\alpha g_{\mu\nu}
    \,,
\end{equation}
which is the typical connection of a \textit{Weyl geometry}. We can conclude that this theory corresponds to Weyl gravity with a metric tensor constrained by Eq.~\eqref{eq:Metric_Weyl}.

Unlike the previous cases, here the secondary breaking down to the Lorentz group is not postulated, but rather a direct consequence of the formalism through a Higgs-like (but Higgs-less) mechanism, as shown in Refs.~\cite{Ghilencea_2019, ghilencea2025quantumgravityweylconformal}. In practice, at the fourth order in the derivative expansion of the EFT the gravity sector is described by two quadratic terms involving the scalar curvature and the Weyl tensor. The $R^2$ term contains an additional spurious scalar degree of freedom that drives the spontaneous symmetry breaking of $SO(3,1) \ltimes \mathbb{R}_{{\rm dil}}$ down to $SO(3,1)$.

In agreement with this discussion, the symmetry breaking scale $\Lambda_{\operatorname{SSB}}$ and the Planck scale $M_p$ are related by
\begin{equation}
    M_p = \frac{\Lambda_{\operatorname{SSB}}}{\sqrt{6}\xi} \,,
\end{equation}
where $\xi$ is the perturbative coupling of the $R^2$ term (see also Ref.~\cite{ghilencea2025quantumgravityweylconformal}).

Applying the coset construction to this symmetry breaking mechanism turns out to affect only the Maurer--Cartan component along the dilatations, which now becomes
\begin{equation}
    \hat{\omega}_{\lambda} = (\partial_\mu \lambda(x) + b_\mu) \dd x^\mu
    \,,
\end{equation}
where $\lambda(x)$ is the (unphysical) Goldstone boson associated with dilatations. It transforms under a Weyl transformation as in Eq.~\eqref{eq:Tranform_dil}, whereas the gauge field $b_\mu$ transforms by construction as
\begin{equation}
    b_\mu \longrightarrow b'_\mu = b_\mu + \partial_\mu \alpha(x)
    \,.
\end{equation}
These two transformation laws suggest that we can define a manifestly invariant object as
\begin{equation}
    E_\mu = b_\mu - \partial_\mu \lambda
    \,,
\end{equation}
which, in the unitary gauge $\alpha(x) = -\lambda(x)$, reduces to $b_\mu$. In this manner, a mass term for the gauge field can be defined; in other words, $b_\mu$ acquires a mass proportional to the Planck mass via a Stueckelberg mechanism. Consequently, the gauge field decouples at low energies, in agreement with the absence of observable nonmetricity effects in standard measurements.

\section{Discussion}

In this work we have discussed General Relativity and some of its modifications as a manifestation of a common physical mechanism, namely the spacetime symmetry breaking of a UV symmetry. The breaking is represented by the Coset $G/H$, in which $G$ is the general linear group and $H$ is any of its subgroups that contains the entire Lorentz group, and we considered four main breaking patterns, of which only one (to our knowledge) was previously discussed more extensively in Ref.~\cite{Tomboulis}.
The four different patterns have led to different effective descriptions of gravity, because they depend on slightly different degrees of freedom,
however all the found descriptions are known to be equivalent to standard GR at the leading order in the derivative expansion.

An important point of this paper is that we noted that the Coset Construction formalism does not provide a unique spin connection, and we needed to introduce 
an arbitrary parameters $c_1$ which can turn on additional geometric structures over spacetime. While our current approach cannot fix this arbitrariness, one can always make an assumption involving the absence of torsion as in Ref.~\cite{Tomboulis},
which leads to the result $c_1 = -1$ described in the main text.
Another possibility, in agreement with the EFT spirit, is to construct the most general effective action at quadratic order and then study its associated dynamics. Two possibilities, not explored in the present work, could be
\begin{itemize}
    \item either to perform a study about the stability of the theory to see whether the tuning $c_1=-1$ is required to avoid ghosts,
    \item or to leave this freedom in the resulting theory and investigate possible consequences. 
\end{itemize}
Concerning the latter, in the presence of torsion, a possible invariant would be
\begin{equation}
    \int T^a \wedge T_a \,,
\end{equation}
that could lead to an explicit violation of parity under spatial inversion. Then one could fix the coefficients requiring parity invariance of the theory or perform a phenomenological matching investigating the possible measured effects of such violation, as done, for example, in Chern-Simons modified gravity \cite{Alexander_2009}.

We conclude by mentioning some possible refinements of our analysis. First of all, the results do not depend on the specifics of the underlying UV theory, but only on the assumption of the existence of a global symmetry group $G_A(4,\mathbb{R})$, so it is important to understand the viability of such an assumption. On the one hand, such a theory could be problematic due to the known difficulties involving the coupling of fermions, because to a known theorem regarding the non-existence of finite-dimensional spinor representations of $GL(4,\mathbb{R})$. This issue is discussed in Ref.~\cite{Lindwasser_2022}, where spinors are introduced through the so-called affine infinite-dimensional spinors, which provide a possible order parameter for the symmetry breaking but also introduce many more spins from the point of view of the Lorentz group.

On the other hand, a potential problem is that, according to the Swampland program, no global symmetry could exist in a quantum gravity theory; see for example Ref.~\cite{harlow2019symmetriesquantumfieldtheory} for a discussion in the AdS/CFT correspondence scenario. In order to be consistent with this, our construction must be modified by considering the group $G_A(4,\mathbb{R})$ as gauged or explicitly broken in the UV theory. While the first case is not practicable for us,\footnote{%
Indeed, if the original $G_A(4,\mathbb{R})$ were gauged, the associated Goldstone boson (i.e., the graviton) would not be a physical state of the theory, and all the constructions of this paper would be invalidated. There are approaches in which the group $GL(4,\mathbb{R})$ is gauged, but they also include the local Lorentz group and provide redundant descriptions of gravity with more gauge symmetries \cite{Floreanini:1989hq,Percacci:2009ij}.
}
the second case can be studied in perfect analogy with chiral effective theory using the spurions method \cite[\S 3.2]{Penco:2020kvy}. In this case, the graviton would be a pseudo-Goldstone boson with a mass suppressed by the symmetry breaking scale. Hence, at low energies this mass becomes negligible and we recover the results discussed above. Nevertheless, it could be interesting to see if the Coset Construction constrains the resulting massive gravity theory, possibly forcing the parameters to constitute a dRG theory. 

A final possibility that we want to mention is that of reconsidering our analysis in relation to generalized symmetries, which have risen in popularity in recent years, and according to which the graviton should be the Goldstone boson of some underlying generalized symmetry. Progress has been made in describing the graviton in this manner at the linearized level, e.g., Ref.~\cite{Hull:2024bcl}, and the graviton is interpreted as the Nambu-Goldstone of a particular shift symmetry and its dual. It would be interesting to discuss whether there is a $GL_A(4,\mathbb{R})$ interpretation of this procedure given that the two approaches aim at similar conclusions, but with rather different methods.

\smallskip

\paragraph*{Acknowledgments and contributions.}
This work is based on the first author's MSc thesis submitted at the University of Pisa and supervised by the second author.
TV is grateful to Stefano Bolognesi for comments and criticism as referee of the original thesis.
OZ is grateful to Roberto Percacci and Dario Sauro for important conversations that inspired this project.

\appendix

\section{Commutation relations of $G_A(4,\mathbb{R})$}\label{app:algebra}

For completeness, we list below the full set of commutation relations for the affine general linear algebra $\mathfrak{g}_A(4,\mathbb{R})$ already given in Eq.~\eqref{GLalgebra} but expressed in terms of the basis that includes the Poincar\'e, shear and dilatation generators.
The Poincar\'e generators satisfy the usual algebra
\begin{equation}
\label{groupalgebra-1}
	\begin{split}
    	[P^a, P^b] &= 0 \,, \\
		[J^{ab}, P^c] &= -i(\eta^{ac}P^b - \eta^{bc}P^a) \,, \\
        [J^{ab}, J^{cd}] &= -i(\eta^{ac}J^{bd} - \eta^{ad}J^{bc} - \eta^{bc}J^{ad} + \eta^{bd}J^{ac}) \,.
    \end{split}
\end{equation}
The complete affine algebra \eqref{GLalgebra} requires the inclusion in the above of the symmetric generators $T_{ab}$ with relations
\begin{equation}
\label{groupalgebra-2}
	\begin{split}
    	[T^{ab}, P^c] &= -i(\eta^{ac}P^b + \eta^{bc}P^a) \,, \\
        [J^{ab}, T^{cd}] &= +i(\eta^{bc}T^{ad} + \eta^{bd}T^{ac} - \eta^{ad}T^{bc} - \eta^{ac}T^{bd}) \,, \\
        [T^{ab}, T^{cd}] &= +i(\eta^{ac}J^{bd} + \eta^{ad}J^{bc} + \eta^{bc}J^{ad} + \eta^{bd}J^{ac}) \,,
    \end{split}
\end{equation}
which are needed for the breaking pattern of Sect.~\ref{sect:Lorentz}.

In order to see the two other subalgebras, we may first include $D=T^a{}_a$, which gives an algebra that we could refer to as ``scaling Poincar\'e algebra'' \cite{Nakayama:2023xzu},
\begin{equation}
\label{groupalgebra-3}
	\begin{split}
    	[D, P^c] = -i P^c \,, &\qquad
        [J^{ab}, D] = 0 \,,
    \end{split}
\end{equation}
and obviously $[D,D]=0$.

Alternatively including in the Poincar\'e algebra only traceless symmetric generators $S_{ab}=T_{ab}-\frac{1}{4}\eta_{ab}D$, we need the relations
\begin{equation}
\label{groupalgebra-4a}
	\begin{split}
    	[S^{ab}, P^c] &= -i\left(\eta^{ac}P^b + \eta^{bc}P^a - \frac{1}{2}\eta^{ab}P^c\right) \,, \\
        [J^{ab}, S^{cd}] &= +i(\eta^{bc}S^{ad} + \eta^{bd}S^{ac} - \eta^{ad}S^{bc} - \eta^{ac}S^{bd}) \,,\\
       [S^{ab}, S^{cd}] &= +i(\eta^{ac}J^{bd} + \eta^{ad}J^{bc} + \eta^{bc}J^{ad} + \eta^{bd}J^{ac}) \,,
    \end{split}
\end{equation}
which give the $SL(4,\mathbb{R})$ algebra.
In the basis $(P,J,S,D)$ the complete affine algebra is recovered by including the final commutator $[S^{ab}, D] = 0$.

\vspace{1cm}



\begin{thebibliography}{99}

\bibitem{CCWZ1}
S.~R.~Coleman, J.~Wess and B.~Zumino,
Phys. Rev. \textbf{177} (1969), 2239-2247

\bibitem{CCWZ2}
C.~G.~Callan, Jr., S.~R.~Coleman, J.~Wess and B.~Zumino,
Phys. Rev. \textbf{177} (1969), 2247-2250

\bibitem{Volkov:1973ix}
D.~V.~Volkov,
Fiz. Elem. Chast. Atom. Yadra \textbf{4} (1973), 3-41

\bibitem{Ogievetsky:1974}
V.~I.~Ogievetsky,
\emph{Nonlinear realizations of internal and space-time symmetries},
in \emph{X-th Winter School of Theoretical Physics in Karpacz},
Acta Universitatis Wratislaviensis, No.\ 207, pp. 117--141,
Wroclaw (1974).

\bibitem{Tomboulis}
E.~T.~Tomboulis,
Phys. Rev. D \textbf{84} (2011), 084018
[arXiv:1105.5848 [hep-th]].


\bibitem{sarmiento2025brokenweylgravitydynamical}
L.~F.~Sarmiento and I.~M.~Rodr{\'\i}guez,
[arXiv:2510.08778 [hep-th]].

\bibitem{Cartan:1981}
E.~Cartan and A.~Mercier, \emph{The Theory of Spinors} (Dover Publications, New York, 1981).
\bibitem{Brauner:2024juy}
T.~Brauner,
Lect. Notes Phys. \textbf{1023} (2024), pp.
2024,
ISBN 978-3-031-48377-6, 978-3-031-48378-3
[arXiv:2404.14518 [hep-th]].

\bibitem{BeltranJimenez:2019esp}
J.~Beltr{\'a}n Jim{\'e}nez, L.~Heisenberg and T.~S.~Koivisto,
Universe \textbf{5} (2019) no.7, 173
doi:10.3390/universe5070173
[arXiv:1903.06830 [hep-th]].

\bibitem{Delacretaz2014}
L.~V.~Delacr{\'e}taz, S.~Endlich, A.~Monin, R.~Penco and F.~Riva,
JHEP \textbf{11} (2014), 008
[arXiv:1405.7384 [hep-th]].

\bibitem{Steenrod:1951}
N. Steenrod, \textit{The Topology of Fibre Bundles} (Princeton University Press, Princeton, 1951).

\bibitem{Gielen:2018}
S.~Gielen, R.~de Le{\'o}n Ard{\'o}n and R.~Percacci,
Class. Quant. Grav. \textbf{35} (2018) no.19, 195009
[arXiv:1805.11626 [gr-qc]].

\bibitem{Codello:2012sn}
A.~Codello, G.~D'Odorico, C.~Pagani and R.~Percacci,
Class. Quant. Grav. \textbf{30} (2013), 115015
[arXiv:1210.3284 [hep-th]].

\bibitem{Sauro:2022chz}
D.~Sauro and O.~Zanusso,
Class. Quant. Grav. \textbf{39} (2022) no.18, 185001
[arXiv:2203.08692 [hep-th]].

\bibitem{Zanusso:2023vkn}
O.~Zanusso,
Phys. Rev. D \textbf{108} (2023) no.12, 125018
[arXiv:2309.09598 [hep-th]].

\bibitem{Ghilencea_2019}
D.~M.~Ghilencea,
JHEP \textbf{03} (2019), 049
[arXiv:1812.08613 [hep-th]].

\bibitem{ghilencea2025quantumgravityweylconformal}
D.~M.~Ghilencea,
Eur. Phys. J. C \textbf{85} (2025) no.7, 815
[arXiv:2408.07160 [hep-th]].

\bibitem{Alexander_2009}
S.~Alexander and N.~Yunes,
Phys. Rept. \textbf{480} (2009), 1-55
[arXiv:0907.2562 [hep-th]].

\bibitem{Lindwasser_2022}
L.~W.~Lindwasser and E.~T.~Tomboulis,
Phys. Rev. D \textbf{106} (2022) no.8, 084026
[arXiv:2207.01067 [hep-th]].

\bibitem{harlow2019symmetriesquantumfieldtheory}
D.~Harlow and H.~Ooguri,
Commun. Math. Phys. \textbf{383} (2021) no.3, 1669-1804
[arXiv:1810.05338 [hep-th]].

\bibitem{Floreanini:1989hq}
R.~Floreanini and R.~Percacci,
Class. Quant. Grav. \textbf{7} (1990), 975

\bibitem{Percacci:2009ij}
R.~Percacci,
PoS \textbf{ISFTG} (2009), 011
[arXiv:0910.5167 [hep-th]].


\bibitem{Penco:2020kvy}
R.~Penco,
[arXiv:2006.16285 [hep-th]].

\bibitem{Hull:2024bcl}
C.~Hull, M.~L.~Hutt and U.~Lindstr{\"o}m,
JHEP \textbf{04} (2025), 046
[arXiv:2409.00178 [hep-th]].

\bibitem{Nakayama:2023xzu}
Y.~Nakayama,
[arXiv:2311.10254 [hep-th]].



\end{thebibliography}
\end{document}